\documentclass[12pt]{article}
\usepackage{latexsym}
\usepackage{amsmath}
\usepackage{indentfirst}
\usepackage{cite} 
\usepackage{graphicx} 

\long\def\ca#1\cb{} 

\newcommand{\ad}{^\dagger }

\newcommand{\AND}{\mbox{\small AND}}

\newcommand{\becs}{\begin{cases}}
\newcommand{\bem}{\begin{matrix}}

\newcommand{\dya}[1]{|#1\rangle\langle#1|}

\newcommand{\encs}{\end{cases}}
\newcommand{\enm}{\end{matrix}}

\newcommand{\ket}[1]{|#1\rangle }

\newcommand{\lra}{\leftrightarrow }

\newcommand{\NOT}{\mbox{\small NOT}}
\newcommand{\od}{\odot }

\newcommand{\ra}{\rightarrow }
\newcommand{\Ra}{\Rightarrow }

\newcommand{\st}{\sqrt{2}}

\newcommand{\PC}{{\mathcal P}}

\newcommand{\al}{\alpha }
\newcommand{\bt}{\beta }

\def\outl#1{\par{\medskip\noindent\hspace*{0.1cm}\bf
      \mathversion{bold}#1\mathversion{normal}\smallskip} }
   \def\xa{} \def\xb{}  

 \def\outl#1{}\def\xa{}\def\xb{}

\ca
 \def\outl#1{\par{\medskip\noindent\hspace*{.5cm}\bf
      \mathversion{bold}#1\mathversion{normal}\smallskip} }
 \long\def\xa#1\xb{} 
  
\cb

\begin{document}

\title{Consistent Quantum Causes}
\author{Robert B. Griffiths\thanks{Electronic address: rgrif@cmu.edu}\\
Department of Physics\\
Carnegie Mellon University\\
Pittsburgh, PA 15213}

\date{Version of 7 December 2024}
\maketitle

\xa
\begin{abstract}
  Developing a quantum analog of the modern classical theory of causation, as
  formulated by Pearl and others using directed acyclic graphs, requires a
  theory of random or stochastic time development at the microscopic level,
  where the noncommutation of Hilbert-space projectors cannot be ignored. The
  Consistent Histories approach provides such a theory. How it works is shown
  by applying it to simple examples involving beam splitters and a Mach-Zehnder
  interferometer. It justifies the usual laboratory intuition that properly
  tested apparatus can reveal the earlier microscopic cause (e.g., as in
  radioactive decay) of a later macroscopic meassurement outcome. This approach
  is further illustrated by how it resolves the Bell inequalities paradox.
  The use of quantum circuits in discussions of quantum information in a
  time-irreversible manner can prevent the proper identification of earlier
  causes; this is illustrated using a specific circuit in the case of Bell
  inequalities. The approach to quantum causes known as Quantum Causal Models
  fails becuase it is not based upon a satisfactory theory of quantum random
  processes.
\end{abstract}

\xb
\tableofcontents
\xa

\xb%
\section{Introduction \label{sct1}} 
\xa%

\xb%
\outl{`Cause' in laboratory \& everyday life. Cause $\lra$ correlation, but
cause precedes effect, and issue of a common cause }%
\xa%

Understanding causes is significant for interpreting laboratory experiments as
well as in everyday life. The mathematical description of causes uses
probability theory; causes are a type of statistical correlation. They are
special in that a cause precedes the corresponding effect in time, and in the
absence of its cause the effect would not have taken place. In addition it is
important to identify cases in which one event precedes another but is not its
cause, as both are consequences of an earlier common cause.

\xb%
\outl{Classical Theory of causes, Pearl \& DAGs, based on Cl stochastic
  processes}%
\xa%

\xb%
\outl{Present article based on consistent approach to Qm stochastic processes}%
\xa%

A very powerful technique for modeling causes and their effects, and
understanding common causes, has been developed in an approach which makes use
of directed acyclic graphs (DAGs). Standard references are \cite{Prl09,SpGs00};
for a compact introduction see Sec. 3 of \cite{Htch18}. It is referred to below
as the \emph{Classical Theory} of causes, since it is widely applied for
discussing macroscopic events and processes. Mathematically it is based on
\emph{classical stochastic processes}, the standard theory of random time
development. Thus one would expect that a quantum counterpart of the Classical
Theory applicable to microscopic causes would be built upon a consistent
approach to \emph{quantum stochastic processes}, and this paper indicates how
to do that.

\xb%
\outl{Problems with previous efforts: Textbooks: Unitary up to measurement}%
\xa%

\xb%
\outl{Physics lab: Microscopic stochasticity illustrated by radioactive decay
occurring randomly and before detection of decay products. Satisfactory theory
of Qm causes must deal with this}%
\xa%

One problem with some previous attempts in this direction is that they start
with textbook quantum theory in which there is unitary, and thus deterministic,
time development starting with some quantum state for a microscopic system, and
randomness only appears when that system is subjected to a macroscopic
measurement.
By contrast, in the physics laboratory many experiments are understood to
involve a random time development at the microscopic level before a measurement
takes place, or in the absence of any measurement. Radioactive decay, for
example, occurs in the absence of a detector, and when a decay product is
detected, the detection event is thought to be caused by the earlier random
decay. In certain cases the precise time of that earlier decay can be estimated
from the detection time. Any satisfactory theory of quantum causes needs to
deal with situations of this type if it is to be of interest to other than
philosophers.

\xb%
\outl{Outline of rest of paper. Sec.~\ref{sct2}: Cl \& Qm stochastic histories.
  Readers not familiar w Cl Theory or CH may wish to first look at examples in
  Sec.~\ref{sct3} }%
\xa%

The remainder of this paper is organized as follows. Section~\ref{sct2a}
indicates how the classical theory of stochastic processes underlies the
Classical Theory of causes. Section~\ref{sct2b} summarizes the Consistent
Histories approach to random quantum time development. Readers not already
conversant with the Classical Theory of causes or Consistent Histories may
prefer to first look at the simple examples in Sec.~\ref{sct3} before reading
the more technical discussions in Sec.~\ref{sct2}. These examples provide a
readily accessible physical picture, discussed using elementary mathematics. of
some of the fundamental difficulties that arise when trying to understand the
nature of causes in the quantum domain.

\xb%
\outl{Qm generalization of Classical Theory of causes: Sec.~\ref{sct4}.
Application to Bell inequalities in Sec.~\ref{sct5} shows how Cl reasoning
$\ra$ results that disagree with experiment }%
\xa%

How to extend these examples to a general theory of quantum causes is indicated
in Sec.~\ref{sct4}. This scheme is then illustrated in Sec.~\ref{sct5} where it
is applied to Bell inequalities, such as those of Clauser et al. \cite{CHSH69},
to show how classical reasoning that ignores quantum noncommutation gives rise
to results that, not surprisingly, disagree with experiment.

\xb%
\outl{Relation of Qm causes to some aspects of Qm info theory and Qm circuits
  is in Sec.~\ref{sct6}}%
\xa%

Some aspects of the relationship between quantum information theory and quantum
causes are discussed in Sec.~\ref{sct6}, in particular in connection with 
the use of quantum circuits, as in the well-known text by Nielsen and Chuang
\cite{NlCh00}. When discussing such circuits it is easy to lose sight of the fact
that at a fundamental level quantum dynamics, both unitary and stochastic, is 
time-reversible. Since identifying the cause of a later event requires
reasoning backwards in time, overlooking the time symmetry of quantum theory
leads to difficulties. 

\xb%
\outl{Wood \& Spekkens, QCM: Sec.~\ref{sct7}; Conclusion in Sec.~\ref{sct8}}%
\xa%

A pioneering attempt by Wood and Spekkens \cite{WdSp15} to embed the Classical
Theory of causes into quantum mechanics ran into serious difficulties (``fine
tuning'') in the case of Bell inequality experiments. This motivated the later
development of a mathematically very complicated approach called \emph{Quantum
  Causal Models} (QCMs) by its developers. Problems with this approach are the
subject of Sec.~\ref{sct7}.

The concluding Sec.~\ref{sct8} provides a brief summary together with some
comments and ideas for further research.


\section{Causes \label{sct2} }

\subsection{Classical Random Causes \label{sct2a}}

\xb%
\outl{Cl Theory based on stochastic processes: History $\lra$ sequence of
  events at successive times. Sample space of histories to which probabilities
  (Kolmogorov) can be assigned. Finite sample space is adequate}%
\xa%

The Classical Theory of causes employs the mathematical structure of
\emph{random} or \emph{stochastic processes}---think of random coin flips or a
Markov process. A single run of a stochastic process, e.g., heads on the first
two tosses and tails on the third, gives rise to a specific \emph{history,} a
collection of \emph{events} occurring at a sequence of successive \emph{times}.
The collection of possible histories constitutes a \emph{sample space} of
mutually-exclusive possibilities in the language of standard (Kolmogorov)
probability theory, assumed throughout the discussion that follows.
\emph{Probabilities} are nonnegative numbers summing to $1$ that are assigned
to the individual histories on the basis of a physical theory, mathematical
model or simply guesswork. Choosing a sample space and assigning probabilities
are distinct matters. Collections of elements from the sample space constitute
the \emph{event algebra}, and the probability of a particular collection is the
sum of the probabilities of the elements it contains. For the following
discussion finite discrete sample spaces suffice.

\xb%
\outl{Directed acyclic graphs (DAGs)}%
\xa%

To discuss causes the Classical Theory employs \emph{directed acyclic
  graphs} (DAGs), where nodes denote individual events at specified times, and
the lines with arrows connecting the nodes indicate possible \emph{conditional
probabilities} of later events conditioned on earlier ones. The arrows indicate
a time ordering from past to future. These conditional probabilities together
with those assigned to initial events (nodes with no incoming arrows) generate
the overall probability distribution for the histories that constitute the
corresponding stochastic process. The reader not already familiar with these
graphs and other concepts in the Classical Theory, such as interventions and
the role of common causes, may find it useful to consult \cite{Htch18}.

\subsection{Quantum Stochastic Processes \label{sct2b}}

\xb%
\outl{Extend Cl to Qm theory requires Hilbert space, Qm stochastic processes}%
\xa%

\xb%
\outl{Textbook QM inadequate. Random events $\lra$ measurements.
  Wavefunction collapse.  $\ra$  measurement problem of Qm founds}%
\xa%

Extending the Classical Theory of causes to the quantum domain, where a quantum
Hilbert space replaces a classical phase space, requires a consistent quantum
theory of stochastic processes with a suitable sample space of histories, and a
rule for assigning them probabilities. 
Constructing such a theory, however, runs into serious difficulties if one is
restricted to ideas found in current quantum textbooks, where the approach
to time development for a closed or isolated system involves a deterministic or
unitary time development---a solution to Schr\"odinger's equation---until the
system undergoes a \emph{measurement} by an external apparatus, at which
point a stochastic process occurs with a mysterious ``collapse of the
wavefunction''. The associated obscurities are part of the infamous
``measurement problem'' of quantum foundations.

\xb%
\outl{Physics lab: random processes at microscopic level preceding measurements
(radioactive decay). Micro causes of measurement outcomes}%
\xa%

In physics laboratories, in contrast to textbooks, random time development at
the microscopic level is a common occurrence. Unstable nuclei decay at
relatively precise, but totally random, instants of time, regardless of whether
a measurement apparatus happens to be nearby. If a decay product is detected
the experimenter thinks that the decay occurred at an earlier time, and in
appropriate circumstances can say how much earlier. Another example is
discussed in Sec.~\ref{sct3a} below. These random microscopic events are then
thought of as \emph{causes} of later macroscopic measurement \emph{outcomes}.

\xb%
\outl{Finite-dimensional Hilbert space suffices.. Projectors $\lra$ Qm
  properties (vN); are Qm counterparts of indicators on a classical phase space.
  Energy of harmonic oscillator an example. Negation, conjunction in terms of
  indicators}%
\xa%

\xb%
\outl{Logic operations using indicators. NOT $P$, $P$ AND $Q$}
\xa

As pointed out by von Neumann, Sec.~III.5 of \cite{vNmn32b}, quantum properties
are represented by \emph{projectors} on a quantum \emph{Hilbert} space. (For
present purposes a finite-dimensional Hilbert space will suffice.) A projector
is a Hermitian operator equal to its square, $P=P\ad =P^2$, and can be
understood as the quantum counterpart of an \emph{indicator}, a function which
on a classical phase space takes the value $1$ where some property---e.g.,
energy $E$ less than some $E_0$ for a harmonic oscillator---is true, and $0$ at
points where it is not true. Certain logical operations can be represented by
indicators (think of Venn diagrams). The negation of property $P$, $\NOT\ P$,
is the indicator $I-P$, where $I$ is the identity indicator taking the value
$1$ everywhere. Conjunction of two properties, $P\ \AND\ Q$, is the indicator
$PQ$, and so forth.

\xb%
\outl{Logic of Qm projectors. Negation $I-P$. Conjunction $PQ=QP$, but product
  in either order not a projector when $PQ\neq QP$}%
\xa%

\xb%
\outl{Noncommutation the Qm-Cl boundary. Qm Logic}%
\xa%

Similarly in the quantum case the negation of $P$ is the projector $I-P$, where
$I$ is the identity operator, and the conjunction ``$P$ and $Q$'' is the
projector $PQ$ \emph{provided} $P$ and $Q$ \emph{commute}: $PQ=QP$. If they do
\emph{not} commute the product in either order is not a projector, and we
arrive at a feature that marks the boundary between classical and quantum
physics: \emph{noncommutation,} the essence of quantum uncertainty relations
involving $\hbar$. Von Neumann was aware of this, and along with
Birkhoff\cite{BrvN36}, invented \emph{Quantum Logic} to make sense of
``$P\ \AND\ Q$'' in the noncommuting case. Alas, Quantum Logic is so
complicated that it has been of little help in understanding quantum mechanics
and resolving its conceptual difficulties.

\xb%
\outl{CH introduced. Commuting projectors a \emph{framework}. PDI defined;
  $\lra$ sample space. Event algebra. Single framework rule. Cl reasoning in
  noncommuting case $\ra$ paradoxes }%
\xa%

The much simpler Consistent Histories%
\footnote{ A compact overview of CH will be found in \cite{Grff24b}, while
  \cite{Grff02c} is a detailed treatment that remains largely up-to-date. For
  CH as a form of quantum logic, see \cite{Grff14} and Ch.~16 of
  \cite{Grff02c}. How CH principles resolve quantum paradoxes is discussed in
  Chs.~20 to 25 of \cite{Grff02c}, and in \cite{Grff17b} and
  \cite{Grff20}.} 
(CH) approach treats ``$P \AND\ Q$'' as \emph{meaningless} in the noncommuting
case, so one never has to discuss its meaning, and limits quantum reasoning to
collections of commuting projectors called \emph{frameworks}. One starts with a
\emph{projective decomposition of the identity} (PDI), the counterpart of a
\emph{sample space} in standard (Kolmogorov) probability theory, a collection
of mutually orthogonal projectors that sum to the identity operator. The
corresponding \emph{event algebra} consists of all the projectors which are
members of, or sums of members, of the PDI. The CH term ``framework'' is used
for either the PDI or the corresponding event algebra. Its \emph{single
  framework rule} states that any sort of logical reasoning must be carried out
using a collection of commuting projectors; reasoning which makes simultaneous
use of noncommuting projectors is invalid. Note that any collection of
commuting projectors generates, via complements and products, a PDI, and thus
an associated framework, so the single framework rule is in this sense not very
restrictive. What is ruled out is the application of classical reasoning to
situations in which two or more projectors do \emph{not} commute, the source of
endless numbers of unresolved quantum paradoxes.

\xb
\outl{Family of histories. Simplest case: single PDI at each time for all of
  the histories; this PDI can depend upon the time. Orthogonality of histories.
  History PDI. Generalization using 'history Hilbert space'}
\xa

The use of PDIs at a single time is easily extended to a collection or
\emph{family} of histories used to discuss quantum stochastic processes in a
manner analogous to the classical case. For many purposes it suffices to
consider families in which the same sequence of times is used for every history
in the family, and in which at a particular time the projectors associated with
the different histories commute with each other, and thus correspond to a
single PDI, though this common PDI may vary from one time to another. Two such
histories are \emph{orthogonal} provided there is at least one time at which
the corresponding events are represented by orthogonal projectors. A collection
of orthogonal history projectors that sum to the \emph{history identity}, a
history in which the event at every time is the identity $I$, constitutes a
\emph{history PDI}, a sample space analogous to that of a classical stochastic
process. There are more complicated situations for which one needs to use a 
\emph{history Hilbert space}%
\footnote{As first pointed out by Isham\cite{Ishm94}. }. %
For example, two histories in a single consistent family may be orthogonal in
the history Hilbert space even though at a particular time the respective
events are represented by noncommuting projectors. For a general formulation
see \cite{Grff24b} and Ch.~10 of \cite{Grff02c}.

\xb%
\outl{Classical unicity vs Qm pluricity}%
\xa%

At this point it is worth noting an important difference between a classical
phase space and a quantum Hilbert space. A single \emph{unique} point is a
classical phase space represents the exact state of the system at a particular
time: all indicators that take the value $1$ at this point represent properties
the system possesses at this time, and those with value $0$ indicate properties
that are not possessed by the system. This is in contrast to the quantum
Hilbert space, where for any nontrivial projector $P$ there will always be
other projectors that do not commute with $P$, and thus cannot be true or false
in the classical sense. Consequently there is never a projector that
characterizes the ``actual'' quantum state of affairs in a way that agrees with
one's classical intuition. Thus intrinsic to Hilbert-space quantum theory is
the possibility that in some situations there might not be a unique way to
describe a physical system at a particular time: quantum \emph{pluricity} in
contrast to classical \emph{unicity}. Examples that illustrate this are found
below in Secs.~\ref{sct3}, \ref{sct5} and \ref{sct6}.

\xb%
\outl{History framework = sample space of histories. Probabilities for closed
  Qm system. Consistency conditions (GMH) illustrated in Secs.~\ref{sct3a},
  \ref{sct3b}. References to fuller discussions. CH, unlike textbooks, can under
  appropriate conditions assign probabilities at intermediate times.}%
\xa%

\xb%
\outl{SFR extended to case of combining history families when probabilities
cannot be defined}%
\xa%

Given a sample space of histories, a \emph{history framework}, the next task is
to assign them probabilities. In the case of a \emph{closed} quantum system,
the unitary time development corresponding to Schr\"odinger's equation can be
used to assign probabilities provided certain \emph{consistency conditions}%
\footnote{While first introduced in earlier work, the correct formulation of
  these conditions is the medium decoherence condition of Gell-Mann and
  Hartle\cite{GMHr93}} %
are satisfied. This represents a generalization, not restricted to measurement
outcomes, of the usual Born rule. The examples below in Secs.~\ref{sct3a} and
\ref{sct3b} illustrate how this works in some simple cases, as will the
discussion of Bell inequalities in Sec.~\ref{sct5}. The reader is referred to
\cite{Grff24b} and Ch.~10 in \cite{Grff02c} for a comprehensive discussion of
the general case. A crucial difference between CH and textbooks is that the
former allows, under appropriate conditions, assigning probabilities to
microscopic situations in a closed system at various \emph{intermediate times}
before a final measurement takes place. A further complication arises in
situations in which two consistent history frameworks can be combined in a
single sample space, a \emph{common refinement}, but for which the consistency
conditions no longer hold for the combination. In this case the single
framework rule is extended to preclude such combinations if one considers them
part of a closed system and wishes to assign probabilities using the extended
Born rule.

\xb%
\outl{Qml measurement apparatus included in closed Qm system. Full Qm
  description not possible, but WQMM shows how micro causes of macro outcomes
  can be inferred. Wavefn collapse and ``agents'' not needed}%
\xa%

In a situation involving macroscopic measurements of a microscopic system, the
measurement apparatus itself, a physical object which obeys fundamental quantum
principles, must be included in an overall closed system to which probabilities
are assigned using quantum principles that apply to physical objects of any
size. Of course it is impossible to construct a complete quantum description of
a macroscopic apparatus, but there is a general approach which seems
satisfactory for retrieving the essential aspects of such a measurement
process. It is discussed in \cite{Grff17b} in a manner which shows how the
later macroscopic outcome (``pointer position'') can be used to infer an
earlier microscopic cause. There is no need to invoke the textbook notion of
wavefunction collapse, which can be quite misleading---see the discussion
in \cite{Grff24b}---or employ the mysterious ``agents'' that appear in various
discussions of quantum foundations.

\xb%
\outl{Extension to open systems}%
\xa%

Frameworks and probabilities can be employed for \emph{open} quantum systems by
the usual device of treating the system and a separate environment as a total
closed system. As discussed above, in suitable situations macroscopic apparatus
can be included as part of the environment. This is part of the more general
issue of providing a quantum basis for classical physics, for which the 
Gell-Mann and Hartle \cite{GMHr93} approach using \emph{quasiclassical}
concepts and frameworks is a plausible first step; also see \cite{Grff24b}.

The representation of quantum stochastic processes using frameworks as defined
above provides an approach to quantum random causes which is in many
respects parallel to that of the Classical Theory. This is illustrated by the 
simple examples in  Sec.~\ref{sct3} preceding a discussion of the 
general case in Sec.~\ref{sct4}



\section{ Simple Examples \label{sct3}} 

\subsection{ Beam Splitters \label{sct3a} }

\begin{figure}[!h]
\begin{center}
\vspace*{-6cm}
\hspace*{-2.7cm}
 \includegraphics[scale=1]{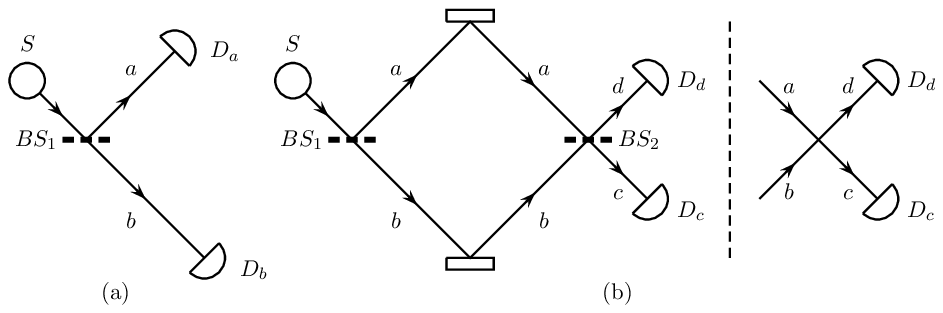}
\vspace*{-6.5cm}
\end{center}
 \vspace*{-12cm}
\caption{(a) Photon source $S$ and beamsplitter $BS_1$ followed by two
  detectors at unequal distances. (b) Add mirrors and a second beamsplitter
  $BS_2$ to form a Mach-Zehnder interferometer. $BS_2$ can be removed, as
  sketched to the right of the dashed line.}
\label{fgr1}
\end{figure}

\xb
\outl{ Figure~\ref{fgr1}(a): Beamsplitter, two detectors. Photon arrival causes
  detection;\\ (1b): Mach-Zehnder with and without 2d beam splitter}%
\xa

\xb%
\outl{ Fig. 1(a). Detection by $D_a \Ra$ particle was on path $a$}%
\xa%

A first gedanken experiment is shown in Fig.~\ref{fgr1}(a). A photon is sent
into a beamsplitter from a source $S$ and emerges by unitary time development
as a coherent superposition
\begin{equation}
\ket{\psi_t} = \al \ket{a_t} + \bt \ket{b_t},
\label{eqn1}
\end{equation}
of two small wave packets traveling along the two macroscopically separated
paths $a$ and $b$, at a time $t$ when the photon has passed through $BS_1$. The
two paths lead to detectors $D_a$ and $D_b$ located at unequal distances from
the beamsplitter. In a particular run of the experiment the photon will be
detected by either $D_a$ or $D_b$, but never by both. Given a large number of
runs, the fraction in which $D_a$ triggers will be approximately $|\al|^2$;
those in which $D_b$ triggers $|\bt|^2$. Someone with experience in the
laboratory will be inclined to think that on a particular run in which $D_a$ is
triggered the photon, after emerging from the beamsplitter, followed path $a$
and was absent from path $b$, while in a run in which $D_b$ detects the photon
it followed path $b$ and was absent from $a$. But since the photon (or other
quantum particle, such as a neutron in an analogous experiment) is invisible,
it would seem useful to do additional tests to corroborate these ideas. Here
are some possibilities.

\xb
\outl{Stochastic splitting of paths at $BS_1$}
\xa

If during certain runs the $a$ path is blocked by some macroscopic absorbing
object---think of this as an intervention---$D_a$ never triggers, while the
fraction of runs in which $D_b$ triggers remains unchanged. Or replace the
block by a mirror which deflects the photon off the $a$ path towards a third
detector $D'_a$. Its average rate is that of $D_a$, whereas the rate of
$D_b$ remains unchanged. These observations support the idea that some sort of
random process occurs at the first beamsplitter $BS_1$, where (to use
anthropomorphic language) the photon randomly chooses either the $a$ or the $b$
path, and then follows it till it reaches the corresponding detector. In
addition, if the photon is emitted at a well-defined time, and the time of
detection measured with sufficient precision, the time to reach detector $D_a$
is less than that required to reach $D_b$, confirming the foregoing intuition.

\xb
\outl{Detection causes wavefn collapse?}
\xa

This seems very odd from a textbook perspective in which random processes only
occur when something is measured, resulting in the collapse of a wavefunction.
Let us explore that idea. Since detector $D_a$ is closer to $BS_1$ than $D_b$,
one might suppose that when the $\al\ket{a_t}$ part of $\ket{\psi_t}$ in
\eqref{eqn1} reaches detector $D_a$ and the photon is detected there, this
instantly removes the amplitude traveling towards $D_b$, which will not be
triggered on this particular run. Such a nonlocal effect seems strange, and is
hard to reconcile with special relativity. An additional difficulty arises in
runs in which $D_a$ does \emph{not} detect the particle. Does such nondetection
result in the amplitude collapsing onto path $b$ at the instant of
nondetection? Or does the later detection by $D_b$ retroactively ``cause'' an
earlier collapse on path $a$ that prevents detection by $D_a$?

\xb
\outl{`Decoherence $\ra$ single path' disproved by interference,
  Fig.~\ref{fgr1}(b)}
\xa

Or might it be that some sort of \emph{decoherence} process, caused by an
interaction of the photon with the environment, either at the beam splitter or
due to gas particles on the way to the detector, is at work, so that the
coherent state \eqref{eqn1} is no longer the correct quantum description
when the two paths have separated by a macroscopic distance? Decoherence
removes quantum interference, so it can be checked with the alternative
experimental arrangement shown in Fig.~\ref{fgr1}(b). Here a second
beamsplitter, $BS_2$, has been added, so that along with the first beamsplitter
$BS_1$ the result is a Mach-Zehnder interferometer. Let us suppose that the
beamsplitters have been chosen in such a way that the photon always emerges on
path $c$, leading to a detection by $D_c$, never $D_d$. Decoherence along the
$a$ or $b$ path inside the interferometer would eliminate the interference, so
can be excluded if the interferometer is functioning properly. 

\xb
\outl{ Compare various situations with $BS_2$ in or out of the way (Wheeler's
  Delayed Choice). Blocking arms. Phase shifters in the two MZI arms }
\xa

Motivated by Wheeler's delayed choice paradox\cite{Whlr78}, let us suppose that
$BS_2$ can be moved out of the way, the situation sketched at the far right of
Fig.~\ref{fgr1}(b), in which case we are back at something that resembles
Fig.~\ref{fgr1}(a), but now the two paths cross. That nothing special happens
when they cross can be checked, as before, by blocking the paths at various
points before they cross, and by moving $D_c$ further away and checking the
timing. Thus one concludes (as did Wheeler) that with $BS_2$ absent the photon
detected by $D_c$ was earlier following path $a$ before reaching the crossing
point, and likewise it was following path $b$ in a run in which it was detected
by $D_d$. Having considered the case with $BS_2$ absent, we must now try and
understand what happens when it is present. One approach is to insert variable
phase shifters in each of the two interferometer arms. While these can change
the ratio of the $D_d$ to $D_c$ counts, it remains fixed if \emph{both} phases
are changed by the \emph{same} amount. Thus the photon must in some sense have
been simultaneously present in \emph{both} arms of the interferometer. This
seems very odd given that with $BS_2$ absent we had good reason to believe the
photon was in either the $a$ arm or the $b$ arm after emerging from $BS_1$. Can
removing $BS_2$, as against leaving it in place, influence what happens
\emph{before} the photon arrives at the crossing point?

\xb%
\outl{Adequate theory of Qm causes should provide explanations. Essentials of
  CH approach in following Sec.~\ref{sct3b} using spin-half particle}%
\xa%

Providing coherent answers to questions of this sort is what one should expect
from an adequate theory of quantum causes. The essentials of the CH approach
and the key respects in which it differs from textbook thinking can best be
explained by an analogous gedanken experiment involving a spin-half particle,
taken up next.



\subsection{ Spin Half \label{sct3b}}

\xb
\outl{Spin-half analog of Fig.~\ref{fgr1}. Spin-half Ag atom on straight path
to SG measuring $S_z$. Paths $a$, $c$ in figure $\lra$ $S_z=+1/2$; $b$, $d$
to $S_z=-1/2$. SG detectors }
\xa

A spin-half analog of the situation in Fig.~\ref{fgr1} for the case of 50-50
beamsplitters allows a mathematical discussion using a two-dimensional Hilbert
space to represent the spin direction, all that is needed for such a particle,
which is simpler than describing the different spatial positions of the photon
in Fig.~\ref{fgr1}. Think of a spin-half silver atom moving along a straight
line until it enters a Stern-Gerlach (SG) measuring device with magnetic field
and field gradient in the $z$ direction, so that it measures $S_z$, yielding
one of the two values $+1/2$ and $-1/2$ in units to $\hbar$. One can think of
$S_z=+1/2$ as the analog of the $a$ path in Fig.~\ref{fgr1}(a), and $S_z=-1/2$
as the analog of the $b$ path, with a single SG detector replacing the pair
$D_a$ and $D_b$.

\xb
\outl{Regions of uniform field replace Fig.~\ref{fgr1} beamsplitters}
\xa
\xb
\outl{Uniform H field can be used to produce particles polarized in any desired
  direction, and to measure $S_w=\pm 1/2$ for any direction $w$ }%
\xa

The beamsplitters in Fig.~\ref{fgr1} are replaced by limited regions of uniform
magnetic field chosen so that as the atom moves through the region its spin
precesses by the desired amount: a unitary transformation on its $d=2$
dimensional Hilbert space. Given a source of spin-polarized atoms, e.g., the
upper beam emerging from a SG apparatus, an appropriate uniform field allows
the preparation of particles with a spin polarization $S_v=+1/2$ for any
spatial direction $v$. Similarly, if the final SG measurement apparatus is
preceded by a suitable region of uniform field, the combination allows the
measurement of any spin component $S_w$, provided passage through the region of
uniform field maps $S_w=+1/2$ to $S_z=+1/2$, and $S_w=-1/2$ to $S_z=-1/2$. That
this analog of $BS_2$ in Fig.~\ref{fgr1}(b) is functioning properly can in
principle be checked by the less convenient process of rotating the final SG
magnet to alter the direction of its magnetic field.

\xb
\outl{Alice prepares $S_x=+1/2$; Bob measures $S_z$. Spin at
  intermediate time =?}
\xa

Suppose that in this way Alice uses the analog of $BS_1$ to prepare a state
with $S_x=+1/2$ which travels on to Bob, whose SG measures $S_z$, the analog of
Fig.~\ref{fgr1}(b) with $BS_2$ absent. The results are random: in roughly half
the runs the outcome corresponds to $S_z=+1/2$, and the other half to
$S_z=-1/2$. What can one say about the spin state during the the time interval
when the particle is moving through the central field-free region on its way
from Alice's preparation to Bob's measurement? Alice might say $S_x=+1/2$,
since she has carried out a lot of tests to check that her apparatus is working
properly, perhaps sending the beam into an SG with field gradient in the $x$
direction. Bob has been equally careful in checking his apparatus, and is sure
that $S_z$ was $-1/2$ in those runs where it indicated this value, and $+1/2$
in the others. But there is no room in the spin-half Hilbert space, no
projector, that can represent the combination ``$S_x=+1/2\ \AND\ S_z=-1/2$'' or
``$S_x=+1/2\ \AND\ S_z=+1/2$''. We seem to have a dispute between two parties.
What shall we to do?

\xb
\outl{Bob on vacation. Alice can analyze preparation/measurement results using
  alternatives $\PC_x$ or $\PC_z$ PDIs an intermediate time.}
\xa

Let us send Bob off on vacation. Alice is perfectly capable of designing and
building both the preparation and the measurement apparatus, and, as with any
competent experimenter, is conversant with the relevant theory. Thus for any
given run she knows both the preparation and the measurement outcome, and faces
the problem of what these data tell her about the spin state at an intermediate
time. She begins with the spin-half history that includes the initial state
$[x+]=\dya{{x+}}$, the projector on the state $S_x=+1/2$, at the
initial time; $[z-]$ at the final time; and the identity operator $I$ at the
intermediate time. This history can be refined in either of two different ways
that will satisfy the consistency conditions, by replacing $I$ at the
intermediate time with one of two PDIs:
\begin{equation}
 \PC_x: \{[x+],[x-]\}; \quad \PC_z: \{[z+], [z-]\},
\label{eqn2}
\end{equation}
where $[\psi]=\dya{\psi}$ is the projector onto the state $\ket{\psi}$, assumed
to be normalized.

\xb%
\outl{Using $\PC_x$ $\ra$ 3-time history family written out explicitly. Implies
$S_x = +1/2$ at intermediate time. Similarly $\PC_z \ra S_z=-1/2$.
SFR means the two cannot be combined }%
\xa%

Using $\PC_x$ yields a family of two possible histories
\begin{equation}
[x+] \od \{[x+],[x-]\} \od [z-],
\label{eqn3}
\end{equation}
of events at three succeisve times: an initial $S_x=+1/2$, a final $S_z=-1/2$,
and different values for $S_x$ at an intermediate time. Using the extended Born
rule means that the probability of $S_x=+1/2$ at the intermediate time,
conditional on the initial and final states, is $1$, and zero for $S_x=-1/2$.
Similarly, using the pair of histories in which the alternatives $\PC_z$ are
employed at the intermediate time leads to the conclusion that $S_z = +1/2$ at
the intermediate time. The single framework rule means that these two families
cannot be combined to yield the nonsensical result that simultaneously
$S_x=+1/2$ \emph{and} $S_z = -1/2$, for which there is no Hilbert-space
projector.

\xb%
\outl{Last minute change in what is measured. Retrocausality mistake. Classical
  analog: color/shape of a piece of paper}%
\xa%

Alice could also decide at the very last instant to measure a different
property, say $S_y$ in place of $S_z$, by altering the detection apparatus.
This ability to change the measurement is sometimes interpreted as having
a \emph{retrocausal} effect: If Alice switches from $S_z$ to $S_y$ it
somehow alters the earlier state of the measured particle. A better perspective
is that Alice's choice does not influence the particle, but simply alters what
she can learn about it by carrying out a measurement. A classical analog would
be deciding between measuring the shape versus the color of a slip of paper.
The difference with the quantum case is that shape and color can be
simultaneously ascribed to the slip of paper, whereas $S_x$ and $S_z$ cannot
simultaneously take on values in the case of a spin-half particle.

\xb
\outl{Omitted from above: Macroscopic measurement apparatus. See CQT, WQMM}
\xa

\xb%
\outl{Alice must \emph{choose} }%
\xa%

While the above discussion identifies the central issue, a number of details
needed for a complete discussion have been omitted. These would include a
proper quantum description, at least in principle, of the measurement apparatus
and its macroscopic outcomes. For these we refer the reader to Chs.~17 and 18
of \cite{Grff02c}, and to \cite{Grff17b} and references given there. The
fundamental point remains the same: Alice, in order to identify the
\emph{cause} of the $S_z$ measurement outcome, must \emph{choose} to use
$\PC_z$ rather than $\PC_x$. There is nothing irrational about this choice, or
similar choices made every day in the laboratory by researchers who regularly
interpret their results in terms of the microscopic causes their apparatus was
designed to detect in situations analogous to those in Fig.~\ref{fgr1}(a).


\section{Consistent Quantum Causes \label{sct4}}

\xb%
\outl{Consistent theory of Qm causes that extends Classical Theory}%
\xa%

\xb%
\outl{Qm stochastic dynamics, Sec.~\ref{sct2b}, uses histories}%
\xa%

\xb%
\outl{Closed Qm system, unitary time development. Extended Born rule assigns
  probabilities to histories in a family satisfying consistency conditions}
\xa%

\xb%
\outl{As in Cl case, conditional probabilities of later given earlier events
  can be related to DAGs}%
\xa%

At this point it is useful to draw together material from Sec.~\ref{sct2} and
from the examples in Sec.~\ref{sct3} to construct a theory of quantum causes
that extends the Classical Theory of Sec.~\ref{sct2a} in a consistent way to the
quantum domain, where it is essential to pay attention to the structure of the
quantum Hilbert space, and in particular to the fact that pairs of projectors
may not commute.
%
This extension requires a consistent formulation of stochastic quantum
dynamics, Sec.~\ref{sct2b}. One starts with a closed quantum system and an
associated set of unitary time development operators generated by
Schr\"odinger's equation. Next choose a family of quantum histories satisfying
the conditions which make possible assigning a consistent set of probabilities
using the extended Born rule. As in the classical case, this overall
distribution can be expressed using a collection of conditional probabilities
for events at a later time in terms of events at an earlier time. Important
features of these conditional probabilities can be expressed using directed
acyclic graphs, as in the Classical Theory.

\xb%
\outl{Measurement apparatus: Use Qcl frameworks of GMH; macro projectors
  commute FAPP, relevant probabilities close to 0 or 1. (Classical chaos
  excluded)}%
\xa%

Measurements carried out using a macroscopic apparatus can be included as part
of the quantum system in the following way. While a detailed microscopic
quantum description of such an apparatus is not possible, its essential
features can be discussed by using a \emph{quasiclassical framework}. The idea,
as proposed by Gell-Mann and Hartle in \cite{GMHr93}, is that classical
mechanics as applied to ordinary macroscopic objects can be regarded as a good
approximation, sufficient for all practical purposes, to an underlying quantum
description that employs a suitable consistent family of histories. Distinct
macroscopic situations, such as the pointer positions in traditional
discussions of measurements, are represented by projectors on enormous
subspaces of the underlying Hilbert space, which commute with each other in an
approximation that is sufficient for all practical purposes. Similarly, the
deterministic time development of Newtonian mechanics is well approximated by a
stochastic evolution in which the relevant probabilities are close to zero or
one. One can make a plausible case for the validity of such an approach as long
as certain obvious exceptions are excluded; e.g., classical chaos with positive
Lyapunov exponents will obviously deviate from quantum theory at small length
scales. It must be admitted that though the ideas in \cite{GMHr93} seem
plausible, they deserve further study.

\xb%
\outl{Macro pointer positions related to micro causes for projective
  measurements, POVMs as in WQMM}%
\xa%

\xb%
\outl{Result: Qm Theory of Causes extends Cl Theory, which is special case of
  Qm Theory}%
\xa%

Given a quasiclassical framework for measurement outcomes the next task is to
relate the pointer positions to preceding causes at the microscopic level where
quantum principles must be applied. The analysis in \cite{Grff17b} indicates in
a general way how this can be done in the case of projective quantum
measurements and POVMs. Combining these ideas makes it plausible that a Quantum
Theory of Causes analogous to the Classical Theory can be constructed using a
consistent set of quantum principles. If that be granted, it follows that the
Classical Theory as applied to ordinary physical processes is consistent with,
and in fact a special case of, the Quantum Theory.

\xb%
\outl{`Interventions' need care in Qm case}%
\xa%

\xb%
\outl{Classical bullet going thru 1 or 2 holes vs Photon in coherent state on
  two paths}%
\xa%

One must, nevertheless, take some care in moving from classical to quantum
theory. For example, \emph{interventions}, such as fixing the value of a
particular variable at a particular time, play an important role in the
Classical Theory. They can be used in the Quantum Theory provided additional
projectors are not introduced in a way that destroys the overall consistency of
the family of histories, something one need not pay attention to in classical
physics. Thus a classical bullet may pass with some probability through one of
two holes in a barrier, and it makes sense to ask how probabilities of later
events are altered if one of the holes is blocked. In a quantum system in which
a single photon can pass in a coherent fashion along two separate paths it can
make a big difference if one of them is blocked, as illustrated in
Sec.~\ref{sct3a} in the case of a Mach-Zehnder interferometer.



\section{ Bell Inequalities\label{sct5}}

\xb%
\outl{CHSH version of BIs violated by lab experiments, in agreement with Qm
  predictions. CH analysis of Bell paradox $\lra$ more complicated situation
  than earlier examples in Sec.~\ref{sct3}. Will use spin-half; easily mapped
  to case of photons}%
\xa%

The reader is no doubt aware that certain Bell inequalities, specifically those
of Clauser-Horne-Shimony-Holt (CHSH) \cite{CHSH69}, are violated by experiments
in precisely the way predicted by quantum theory. The use of consistent quantum
causes for understanding what is wrong with derivations of what might be called
the \emph{Bell paradox} will serve to illustrate their use in a situation more
complicated than those addressed earlier in Secs.~\ref{sct3a} and B.
The discussion below uses spin-half particle language, which is
easily mapped onto the case of entangled photons as measured in the laboratory.

\xb%
\outl{Each round of ``Bell experiment'': Alice and Bob measure one of two
  orthogonal spin components. So 4 different, mutually incompatible, types of
  measurement. Data from different types cannot be combined; SFR of CH. But
  combining is behind CHSH inequalities}%
\xa%

Each round of a Bell experiment consists of measurements by two parties, Alice
and Bob, who are distant from each other, with each making a choice to carry
out one of two possible incompatible measurements, e.g., $S_x$ and $S_z$, that
correspond to mutually perpendicular directions on the Bloch sphere. These
measurements are carried out on an entangled state of the two spin-half
particles, prepared at an earlier time. Hence there are four different, mutually
incompatible, types of measurement carried out jointly by the two parties in
different runs of the experiment, and the statistical data must be accumulated
and analyzed separately for each type of run. The single framework rule of CH
states that these four data sets cannot be combined in the way routinely
used when deriving CHSH inequalities and similar results, as it employs a
process of classical reasoning that ignores the noncommutation of quantum
projectors, and hence the resulting inequalities are invalid.

\xb%
\outl{Mistake for Alice to assume  $S_x$ has an unmeasured value in run where
  $S_z$ was measured. But she is correct in employing \emph{local causality},
contrary to Bell }%
\xa%

Consider in particular a run in which Alice measures $S_z$. It would be a
mistake for her to assume that in \emph{this} run the spin-half particle
\emph{also} possessed a value of $S_x$ that was somehow ``there'', even though
it was not measured. See the remarks in Sec.~\ref{sct3b}. On the other hand,
as she has carried out a projective measurement having properly calibrated her
apparatus, she is correct in concluding that the $S_z$ value revealed by the
(macroscopic) outcome was the value possessed by the spin-half particle just
before the measurement took place: an instance of \emph{local causality} of the
sort that Bell thought inconsistent with quantum theory.

\xb%
\outl{Tracing properties just before measurements back in time $\ra$
  \emph{common cause} of Alice-Bob correlations. Wavefn collapse not needed}%
\xa%

In addition, given that the particle did not pass through a magnetic field on
the way to her apparatus, she can conclude that the spin had the same value at
an earlier time just after the initial entangled state of the two particles was
prepared, and the partner particle started on its way to Bob. For the same
reason Bob can conclude that his measurement reveals the corresponding spin
state of his particle at this earlier time. Thus, by use of an appropriate
family of histories, the observed correlations can be traced backwards in time
to a \emph{common cause}. Alice and Bob can carry out their measurements on a
particular pair of spin-half paticles at the same time or one measurement may
be later than the other: it makes no difference. There is no need to invoke a
mysterious and unphysical ``wavefunction collapse'' in order to understand the
correlations.

\xb%
\outl{Alice, Bob both free to choose what to measure just before arrival of
  particles, or agree upon a list in advance. Much literature on exptl tests of
  Bell inequalities is based on Cl reasoning, so irrelevant. See details in
  \cite{Grff20} }

Note that Alice and Bob are free to choose, at any time up to just before their
respective measurements take place, which spin component to measure, e.g., the
choice could be determined using a random number generator. Or they may employ
a shared list agreed upon in advance as to what choice to make on each run. A
large part of the vast literature on experimental tests of Bell inequalities
becomes irrelevant when classical reasoning is abandoned in favor of
Hilbert-space quantum mechanics, and quantum histories are used to represent
quantum stochastic processes. For further comments and details see
\cite{Grff20}.

\xb%
\outl{CH gets rid of spurious nonlocal influences that violate no-signaling}%
\xa%

Among other things the CH analysis gets rid of nonlocality claims that go back
to Bell: The notion that in the quantum world there are mysterious nonlocal
influences giving rise to statistical correlations seemingly in violations of
special relativity. The simplest explanation of why such influences are never
observed in the laboratory, where experiments have confirmed the
``no-signaling'' principle, is that they do not exist, and are hence incapable
of carrying information from one place to another.



\section{ Quantum Information and Quantum Circuits \label{sct6}}

\subsection{ Introduction \label{sct6a}}
        
\xb
\outl{Classical information theory employs Cl Theory of Causes}
\xa

Information theory deals with the transmission of signals from the past to the
future, as in how the output of a noisy communication channel is related to the
input. Hence an analysis of causes and their effects---the relationship is
typically probabilistic, not deterministic---is necessarily a central aspect
of both the classical and the quantum theory of information. In the classical
case the mathematical foundation is the Classical Theory of Causes,
Sec.~\ref{sct2a}, so in trying to understand quantum information it is natural
to begin with its quantum counterpart. This, however, can run into certain
difficulties,

\xb
\outl{Qm Circuits are typically irreversible. This can create difficulties
when seeking to identify Qm causes}
\xa

The well-known text of Nielsen and Chuang (N\&C)\cite{NlCh00} makes
considerable use of \emph{quantum circuits}. Such a circuit is represented by a
graph in which the nodes represent quantum states at successive times,
connected by lines indicating some sort of time development. This can either be
deterministic, corresponding to a unitary map, or stochastic, such as
represented by a completely positive quantum operator. The latter can be
thought of as a unitary map acting on the system of interest and an
environment, with the latter not directly represented in the quantum circuit.
This is satisfactory as long as the circuit is being used to assign
probabilities to some future possibilities based on some initial data. But
quantum circuits used in this way can prove inadequate when searching for an
earlier quantum cause of a later property or measurement outcome.
 
\xb
\outl{Discrete time steps in Qm circuits $\lra$ discrete CH histories}
\xa

\xb \outl{Key CH $\lra$ N\&C difference illustrated by Griffiths-Niu
  identifying earlier causes in Shor factorization, leading to a more practical
  procedure, vs.\ N\&C turning Griffiths-Niu into principle of 'deferred
  measurement'.} \xa

The discrete time steps in a quantum circuit are analogous to the discrete time
steps in a family of quantum histories. However, a key difference between the
CH and N\&C approaches is associated with the latter's use of a ``principle of
deferred measurement'' in Sec.~4.4 of \cite{NlCh00}. It is, as the authors
state, motivated by an earlier publication by Griffiths and Niu \cite{GrNi96},
in which it was pointed out that the final steps in the famous Shor
factorization algorithm can be made more practical by replacing a final
measurement process on a collection of qubits with a sequence of prior
measurements on individual qubits at succeeding time steps, in which the choice
of a later measurement is determined by the outcomes of earlier measurements.
The resulting simplification continues to be employed in some more recent
discussions of the Shor algorithm, and analogous ideas are found in other
proposals to make intermediate-time measurements part of a quantum process.

\xb \outl{ Griffiths-Niu was a CONCEPTUAL advance, not mere calculational
  procedure} \xa

The basic point relevant to the current discussion is that by identifying
earlier causes the approach in \cite{GrNi96} led to a conceptual advance, which
was lost when N\&C identified ``deferred measurement'' as nothing but a
calculational tool: What could be measured earlier could be put off until a
later measurement. This is quite true, but in replacing a conceptual tool with
a calculation principle N\&C abandoned the advantages of the former. A similar
problem arises in various discussions of causality in relation to Bell
inequalities, which is discussed next.



\subsection{ Quantum Circuit for the Bell Paradox \label{sct6b}}

\xb
\outl{4-qubit Qm circuit for Bell stuff}
\xa

\begin{figure}[h!]
\vspace*{-6cm}
\begin{center}
\hspace*{-9cm}
\includegraphics[scale=1.05]{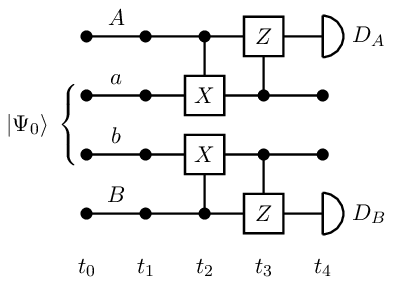}
\vspace*{-19cm}
\end{center}
\caption{ Quantum circuit for a Bell test. The qubits $A$ and $a$ are in Alice's
  domain, while $B$ and $b$ are some distance away in Bob's laboratory, and
  hence there is no interaction between the $A,a$ pair and the $B,b$ pair. The
  $X$ and $Z$ gates are bit-flip and phase-flip, respectively. See the
  discussion in the main text. }
\label{fgr2}
\end{figure}

\xb
\outl{How Fig.~\ref{fgr2} circuit works. Alice chooses initial $A$ $\ra$
  which measurement of $a$ }
\xa

A simple quantum circuit that can be used to understand the role of causes in
the Bell paradox, Sec.~\ref{sct5}, is shown in Fig. 2. By setting the initial
value of qubit $A$ equal to $\ket{0}$ at time $t_0$, Alice's circuit measures
the initial value of $S_{az}$, eigenstates $\ket{0}$ or $\ket{1}$, of qubit
$a$, with eigenvalues of $+\hbar/2$ and $-\hbar/2$, respectively, leading to
the $D_A$ detector output $\pm 1$ at time $t_4$. If, on the other hand, she
sets the initial value of $A$ to $\ket{1}$, the $D_A$ outcomes $\pm 1$
correspond to initial states $\ket{+}$ and $\ket{-}$, eigenstates of $S_{ax}$,
for qubit $a$ time $t_0$, where
\begin{equation}
 \ket{+} = (\ket{0} + \ket{1})/\st,\quad \ket{-} = (\ket{0} - \ket{1})/\st.
\label{eqn6.1}
\end{equation}
The same comments apply to Bob's half of the circuit with qubits $B$ and $b$.
\xb
\outl{``Trivial''  unitaries from $t_0$ to $t_1$, $t_1$ to $t_2$ suggest
  $t_1$ nodes can be omitted}
\xa

Notice that trivial unitary transformations---identity operators---link the
nodes at $t_1$ to their earlier and later counterparts at $t_0$ and $t_2$.
Consequently, if one simply wants to calculate the end result of the overall
unitary transformation from $t_0$ to $t_4$ the $t_1$ nodes can simply be
omitted and replaced by trivial unitaries, straight lines, connecting the $t_0$
to the $t_2$ nodes. But this removes the possibility of using the circuit to
identify the causes at time $t_1$ of the later measurement outcomes.

\xb%
\outl{Alice half of Fig.~\ref{fgr2} $\lra$ her analysis in Sec.~\ref{sct3b}}%
\xa%

The Alice half of the circuit in Fig. 2 corresponds to her analysis of the
single qubit situation in Sec.~\ref{sct3b}, which in turn corresponds to to
that in Fig.~\ref{fgr1} (a). In the laboratory it is very natural to ascribe a
detection in $D_a$ or $D_b$ to a photon having arriving along the corresponding
path. It is this sort of laboratory intuition which is all too easily
overlooked by theoreticians who base their analyses on quantum circuits without
considering their physical counterparts.

\xb%
\outl{Entangled state $\ket{\psi_0}$ for $a$ and $b$ at $t_0$ $\lra$ Bell
  paradox in Sec.~\ref{sct5}}%
\xa%

Next assume that the initial state of qubits $a$ and $b$ at time $t_0$ is
the entangled state
\begin{equation}
 \ket{\Psi_0} = \{ \ket{00}_{ab} - \ket{11}_{ab} \}/\st,
\label{eqn}
\end{equation}
corresponding to the Bell paradox of Sec.~\ref{sct5}. Then each pair ($D_A$,
$D_B$) of measurement outcomes can be associated with the corresponding prior
state of the qubits at time $t_1$, and as these prior states are not mutually
orthogonal it becomes obvious that the results from the four distinct kinds of
experiment cannot be combined because of quantum incompatibility, and thus the
Bell paradox is resolved. Its resolution depends on using the Griffiths-Niu
perspective \cite{GrNi96}, which is easily overlooked in a straightforward but
limited analysis employing quantum circuits.




\section{ Quantum Causal Models \label{sct7}}


\xb%
\outl{Wood-Spekkens difficulties (fine tuning) analyzing Bell
  inequalities. They lacked Qm stochastic processes in contrast to CH, and
  used textbook unitary time development until measurement}%
\xa%

The pioneering attempt of Wood and Spekkens \cite{WdSp15} to produce a quantum
counterpart of the classical approach to causes, one useful for discussing the
Bell inequalities problem, uncovered difficulties in the form of an unnatural
``fine tuning'' condition; see the original paper for details. From the
perspective of the present article the actual problem was the lack of a proper
theory of quantum stochastic processes, and thus a reliance on the
unsatisfactory textbook approach of unitary, and thus deterministic, time
development until a measurement occurs. But that stands in the way of
understanding measurements as potentially revealing a prior microscopic
property. An approach which cannot properly handle the simple situation
discussed in Sec.~\ref{sct3a} can hardly be expected to work in the more
complicated Bell scenario, see Sec.~\ref{sct5}, that involves measuring four
incompatible quantum observables in four separate experimental runs.

\xb%
\outl{Quantum Causal Models (QCMs)}%
\xa%

\xb%
\outl{QCM approach is complicated (`sledgehammer') \& irreversible in time}%
\xa%

Attempts to remedy the problem identified by Wood and Spekkens, and extend
their analysis to more complicated situations, gave rise to an approach called
\emph{Quantum Causal Models} (QCM) by its advocates. There are by now a large
number of papers plus additional preprints devoted to this topic; see the
overview in the recent \cite{Ynao24}. From a CH perspective a fundamental
problem with the QCM approach is that it is built upon the same sort of
thinking that in quantum information studies employs quantum circuits in a way
that is fundamentally irreversible in time. See the discussion in
Sec.~\ref{sct6}. It is bound to fail if a proper analysis of quantum causes
requires a time-reversible mathematical structure analogous to that
used in the classical case. Employing more and more complicated
mathematics---one author \cite{Shrp19} aptly described this as a
``sledgehammer''---only prevents work of this sort from leading to genuine
physical insights, such as those associated with the simple laboratory
situations discussed in Sec.~\ref{sct3}.

\xb%
\outl{No reason why these systems cannot be analyzed in time-reversible way}%
\xa%

There is no obvious reason why a time-reversible approach cannot be used to
analyze some of the more complicated situations that have been studied from the
QCM perspective, and doing so might well give rise to interesting physical
insights as well as, one would hope, simpler and more easily understandable
mathematics.


\section{  'Summary and Conclusion \label{sct8}}

\xb%
\outl{CH $\ra$ framework for Qm stochastic processes without needing
  measurements}%
\xa%

\xb%
\outl{Key ideas: Projectors represent physical properties. Pay attention to
  commutation of projectors. Consistency conditions $\lra$ probabilities of
events related to unitary time development. Physical measurements $\lra$
apparatus described in Qml terms as part of a closed system. CH has no
`measurement problem'}%
\xa%

The consistent histories use of quantum histories based on the mathematical
structure of Hilbert space quantum mechanics provides a framework for
discussing random time development, thus quantum stochastic processes, without
the need to invoke measurements. Its key principles include the use of
projectors on the quantum Hilbert space to represent physical properties,
inference rules that pay strict attention to whether projectors do or do not
commute with one another, and consistency conditions which single out cases in
which probabilities of events in a closed quantum system can be related to the
corresponding unitary time development. Unlike textbook quantum theory,
measurements are not a central concept in the CH approach. Physical
measurements are carried out by physical apparatus which can, at least in
principle, be described in quantum terms provided the measurement
apparatus is included in the total closed system under consideration. Thus the
CH approach has no measurement problem; stated another way, it has resolved the
measurement problems which beset many studies of quantum foundations.

\xb%
\outl{This allows Cl theory of causation to be embedded in Qm domain as per
Sec.~\ref{sct2}. Qm framework $\lra$ Cl sample space. Probs well defined
on the associated event algebra; satisfy the usual rules; consistent with
laboratory practice.
}%
\xa%

\xb%
\outl{Formulation of Qm info using Qm circuits can mislead by concealing the
  fact that the underlying Qm theory is time-reversible, Sec.~\ref{sct6}}%
\xa%

The use of consistent histories provides an immediate way in which to embed the
modern Classical Theory of causation in the quantum domain. In Sec.~\ref{sct2}
it is shown that a quantum framework in which the consistency conditions are
satisfied is the analog of a classical sample space used when discussing
classical stochastic processes. Probabilities for the corresponding event
algebra are then well-defined and satisfy the usual rules. In particular this
allows a discussion of microscopic causes in a way consistent with the way they
are commonly understood in the laboratory.
As noted in Sec.~\ref{sct6}, a formulation of quantum information theory using
quantum circuits can easily overlook the fact that the underlying quantum
theory is fundamentally time-reversible. That feature is important when
analyzing quantum causes, since one is typically carrying out an inference to
events at an earlier time on the basis of later information.

\xb%
\outl{Items that deserve further study}%
\xa%

There remain a number of open problems and issues that deserve further study,
among them the following.

\xb%
\outl{Role of DAGs in the Qm theory}%
\xa%

In the Classical Theory the use of directed acyclic graphs (DAGs) is useful in
many ways, such as providing an intuitive picture of causes, and restricting
the kind of probability distribution compatible with the graph. In the Quantum
Theory it is possible to construct a DAG corresponding to a specific family of
histories, but it is not obvious that this will be a useful idea. Studying this
question using specific examples seems worthwhile.

\xb%
\outl{Relaxing exact CH mathematical conditions}%
\xa%

The consistent histories approach employs a set of exact mathematical
conditions in defining an acceptable family. However, the theoretical (in
contrast to the mathematical) physicist often finds it useful to make
approximations. Suppose that a family of histories is made consistent provided
at some point in time $S_z$ for a spin-half particle is replaced by $S_w$,
where $w$ is a direction in space close to the $z$ axis. Isn't this good
enough? It was pointed out some time ago by Dowker and Kent \cite{DwKn96} that
if the violation of consistency conditions is small there is a `nearby' family
of histories in which they are satisfied exactly. Pursuing this topic further
seems worthwhile, and could conceivably be relevant to studies of quantum
information.

\xb%
\outl{Open systems: Use quantum environment as part of overall closed system}%
\xa%

As mentioned in Sec.~\ref{sct2b}, open quantum systems can be discussed in the
usual way by introducing an environment which, together with the main system of
interest, constitutes a single closed system. Distinguishing the environment
and the main system can be done by using appropriate history families. This
approach is worth pursuing in the hope that it will allow things like
decoherence and Markovian behavior to be studied in a consistent manner with
less arm waving.

\xb%
\outl{Re-examination of situations discussed using QCMs}%
\xa%

A number of situations have been examined from the perspectives of Quantum
Causal Models \cite{Ynao24}, and studying them using tools compatible with the
quantum Hilbert space might provide greater clarity and some interesting
physical insights.

\xb
\section*{Acknowledgements}
\xa

The author is indebted to anonymous reviewers of earlier versions for critical
remarks that have led to substantial revisions and improvements in the
present text. And he is grateful to Carnegie-Mellon University and its Physics
Department for continuing support of his activities as an emeritus faculty
member.


\bibliographystyle{unsrt}


\end{document}